\documentclass[%
onecolumn,%
oneside,%
floats,%
aps,%
prd,%
nobibnotes,
nofootinbib,%
amsmath,%
amssymb,%
amsfonts,%
superscriptaddress,%
]{revtex4}

\usepackage[utf8]{inputenc}
\usepackage{graphicx,array,dcolumn,xcolor}
\usepackage[paperwidth=210mm,paperheight=297mm,centering,hmargin=1.8cm,vmargin=2.5cm]{geometry}
\usepackage[hidelinks]{hyperref}

\PassOptionsToPackage{square,numbers,%
sort&compress,%
}{natbib}
\usepackage{natbib}
\newcommand{\be}{\begin{equation}}
\newcommand{\ee}{\end{equation}}

\def\be{\begin{equation}}
\def\ee{\end{equation}}
\def\mn{{\mu\nu}}

\renewcommand\rho{\varrho}
\renewcommand\tilde{\widetilde}

\graphicspath{{./arXiv/}}

\begin{document}

\title{Cosmic ray production in modified gravity}

\author{E.V. Arbuzova}
\email{arbuzova@uni-dubna.ru}
\affiliation{Novosibirsk State University, Novosibirsk, 630090, Russia}
\affiliation{Department of Higher Mathematics, Dubna State University, 141980 Dubna, Russia}

\author{A.D. Dolgov}
\email{dolgov@fe.infn.it}
\affiliation{Novosibirsk State University, Novosibirsk, 630090, Russia}
\affiliation{ITEP, Bol. Cheremushkinsaya ul., 25, 113259 Moscow, Russia}

\author{L. Reverberi}
\email{lorenzo.reverberi@fzu.cz}
\affiliation{%
Central European Institute for Cosmology and Fundamental Physics (CEICO),\\
Fyzik\'{a}ln\'{\i} ust\'{a}v Akademie v\v{e}d \v{C}R, Na Slovance 2, 182 21 Praha 8, Czech Republic}


\begin{abstract}
This paper is a reply to the criticism of our work on particle production in modified gravity by D. Gorbunov and A. Tokareva. We show that 
their arguments against efficient particle production are invalid. $F(R)$ theories can lead to an efficient generation of high energy cosmic 
rays in contracting systems.
\end{abstract}

\maketitle

\section{Introduction}
\label{s-intro}

In our paper~\cite{Arbuzova:2010iu} we have found that in some versions of $F(R)$-theories curvature scalar $R$ reaches infinity in objects 
with a rising mass/energy density, as first predicted in~\cite{Frolov:2008uf} and further investigated in~\cite{Reverberi:2012ew}. The singularity 
can be avoided if 
an $R^2$-term~\cite{Starobinsky:1980te} is added to the action. In this case $R$ remains finite but reaches very large values in the process of high frequency oscillations. 

{Imposing these arguments one should keep in mind that both General Relativity (GR) and F(R)-modified theories are not fundamental but
effective theories valid in a low energy limit. In particular GR is broken and looses its predictability at the energies above the Planck  
energy $E_{Pl} = m_{Pl} \approx 10^{19}$ GeV or when the curvature scalar reaches the value of the order $m_{Pl}^2$. 
The range of applicability of $F(R)$-theories depends upon the concrete form of the function $F(R)$. In particularly, for 
$F(R) = R^2/6 m^2$ (the case which is discussed below) the ultraviolet cut-off may be around $R \sim m_{Pl} m$. As it is known,
the addition of $R^2$-term into the gravitational action prevents rising of $R$ to infinitely large values. It can be verified 
(see below, the end of sec.~\ref{ss-scal-pot}) that the stabilization is achieved at $R\ll m_{Pl} m $ and even at $R \ll m^2$. 
So the theory is stabilized below its intrinsic ultraviolet cut-off.

One more problem of $F(R)$ theories may emerge due to violation of the energy condition. This issue is analyzed in 
ref~\cite{Appleby:2009uf}, where the properties of the theory are established which permit to overcome this possible difficulty.
}

In  our papers~\cite{Arbuzova:2012su,Arbuzova:2013ina} we argued that such high frequency oscillations could lead to an efficient production of energetic particles, which may in principle give a noticeable contribution to the flux of energetic cosmic rays. These results have been recently criticized by Gorbunov and Tokareva (GT from here on) in Refs.~\cite{Gorbunov:2014fwa,Gorbunov:2014eda}, where the authors claim that the flux of the produced particles is much lower than our prediction, and essentially negligible. In this note we analyze the criticism by GT and show that the arguments against our results are erroneous or inapplicable.

The main points of the GT criticism of our results include the following:
\begin{enumerate}
 \item The choice of the initial conditions in the differential equation describing the evolution of $R(t)$.
 \item The magnitude of energy supply to the oscillations of curvature.
 \item A smaller probability, in comparison with our estimates, of particle production.
 \item An effect of the scalaron decay on the amplitude of $R(t)$.
 \item The impact  of discretization of matter.
\end{enumerate}

Our reply-paper is organized as follows. In Sec.~\ref{s-model} we briefly summarise the content of our 
works~\cite{Arbuzova:2012su,Arbuzova:2013ina}, introduce notations, and present the main equations and results. In Sec.~\ref{s-init-cond} we discuss the problem of the initial conditions and argue that the conditions which we have chosen in our works as the natural ones can be rigorously obtained from the study of the curvature evolution at the onset of the cosmological structure formation. Though, strictly speaking, {the choice of the initial conditions in our works~\cite{Arbuzova:2012su,Arbuzova:2013ina} 
}
was not rigorously justified, the treatment presented in Sec.~\ref{s-init-cond} proves that 
{these conditions have been guessed} correctly.
In Sec. \ref{s-class-quan} the process of particle production in contemporary universe {by} 
collapsing objects is compared with particle production
{by cosmological background} in the early universe. It is shown that the former is much more 
efficient. Section \ref{s-non-harm} is dedicated to an important effects
of non-harmonicity and deviation { from } the adiabaticity for stimulation of particle production. In Sec.~\ref{s-discrete} a possible impact
of the discreteness of matter is considered  and it is found there that such effects do not lead to suppression of the $R$-oscillations.
In Section~\ref{s-conclusion} we summarize our counterarguments and conclude.

\section{The model and basic equations}
\label{s-model}

In this section we briefly present the main features and notations of the papers~\cite{Arbuzova:2012su,Arbuzova:2013ina} to make the presentation self-contained. For more details we {address} the reader to the quoted works.

\subsection{Modified gravity without \texorpdfstring{$R^2$}{R2} term}
\label{ss-no-R}

We consider the model with the action:
\be
S = \frac{m_{Pl}^2}{16\pi} \int d^4 x \sqrt{-g} \left[ R + F(R)\right]+S_m, 
\label{A1}
\ee 
where $m_{Pl} = G_N^{-1/2} \simeq 1.22 \times 10^{19}$ GeV is the Planck mass, $R$ is the scalar curvature, and $S_m$ is the matter action. The first term in Eq.~(\ref{A1}) is the usual General Relativity (Einstein-Hilbert) action and $F(R)$ is a non-linear function 
of the scalar curvature. In this subsection, {following Refs.~\cite{Arbuzova:2012su,Arbuzova:2013ina}}, 
we take $F(R)$ equal to:
\be
\label{eq:model-0}
F(R) = -\lambda R_c\left[1-\left(1+\frac{R^2}{R_c^2}\right)^{-n}\right]\,. 
\ee
Here $n$ is an integer, $\lambda = {\rm const.} > 0$ (in what follows we take $\lambda =1$), and $| R_c |$ is of the order of $8\pi \rho_c / m_{Pl}^2$, where $\rho_c$ is the present day value of the total cosmological energy density, so $R_c \sim 1/t_U^2$, where $t_U \approx 4\times 10^{17}$s is the universe age.

This model is not entirely realistic because it requires a past-time singularity~\cite{Appleby:2009uf} and, which is even worse, in systems with rising energy density it can lead to a singularity in a finite time in the future~\cite{Frolov:2008uf,Arbuzova:2010iu,Reverberi:2012ew}. Still, investigating this model is instructive, because it is technically much simpler than a more realistic model where the singularities are eliminated by the $R^2$ term added to the action, see below Eq.~(\ref{eq:model}). Many essential features of the realistic scenario, particularly at small curvatures, can be understood with this simpler model.
In what follows we consider objects with $|R|\gg |R_c|$.

The equations of motion which follow from the action (\ref{A1}) have the form:
\be
\left( 1 + F'_{R}\right) R_{\mu\nu} -\frac{1}{2}\left( R + F\right)g_{\mu\nu}
+ \left( g_{\mu\nu} D_\alpha D^\alpha - D_\mu D_\nu \right) F'_{R} = 
\frac{8\pi T^{(m)}_{\mu\nu}}{m_{Pl}^2}\,,
\label{eq-of-mot}
\ee 
where $F'_{R}= dF/dR$, $D_\mu$ is the covariant derivative, and $T^{(m)}_{\mu\nu}$ is the energy-momentum tensor of matter. Taking the trace of Eq.~(\ref{eq-of-mot}) leads to the equation:
\be
3 D^2 F'_R -R + R F'_R - 2F = 8\pi T_\mu^\mu /m_{Pl}^2\,.
\label{trace-0}
\ee
This is a closed equation for $R$ except for metric depending terms in the covariant derivative and in $T^\mu_\mu$. If the metric slightly 
differs from the flat Minkowski
one, equation (\ref{trace-0}) would contain only one unknown scalar function which completely determines the evolution of $R$. In this limit the equation can be reduced to the simple oscillator form:
\be
(\partial^2_t - \Delta) w + U'(w) = 0
\label{eq-for-w}
\ee
for the function
 \be
w \equiv - F'_R \approx 2n\lambda \left(\frac{R_c}{R}\right)^{2n+1} \,,
\label{F'-of-R}
\ee
where the potential $U(w)$ is defined as:
\be
U(w) = \frac{1}{3}\left( \tilde T - 2\lambda R_c\right) w + 
\frac{R_c}{3} \left[ \frac{q^\nu}{2n\nu} w^{2n\nu}+ \left(q^\nu
+\frac{2\lambda}{q^{2n\nu} } \right) \,\frac{w^{1+2n\nu}}{1+2n\nu}\right]\,,
\label{U-of-w}
\ee
with $\nu = 1/(2n+1)$, $q= 2n\lambda$, $U'(w)=dU/dw$, and $\tilde T = 8\pi T_\mu^\mu /m_{Pl}^2$. Moreover, we are considering the case $|R| \gg R_c$, since in realistic astrophysical systems $\tilde T\gg R_c$. Their ratio is about $\tilde T/R_c \sim \rho_m/\rho_c \gg 1$ and hence $w\ll 1$. Thus the first term in square brackets in Eq.~(\ref{U-of-w}) dominates. The potential $U$ would depend on time if the mass
density of the object under scrutiny also changes with time, $\tilde T=\tilde T(t)$.

If only the dominant terms are kept in equations (\ref{eq-for-w}) and (\ref{U-of-w}) and if the space derivatives are neglected, the equation (\ref{eq-for-w}) simplifies to:
\be
\ddot w + \tilde T/3 - \frac{q^\nu (-R_c)}{3w^\nu}=0\,.
\label{eq-w-simple}
\ee 
It is convenient to introduce the dimensionless quantities:
\be
t = \gamma \tau,\,\,\, w = \beta \zeta\,,
\label{dimless}
\ee
where $\beta$ and $\gamma$ are so chosen (see below) that the equation for {$\zeta$} becomes particularly simple:
\be
\zeta'' - \zeta^{-\nu} + (1+\kappa \tau) = 0\,.
\label{eq-for-z}
\ee
Here a prime denotes differentiation with respect to $\tau$ and the trace
of the energy-momentum tensor of matter is parametrised as:
\be
\tilde T(t) = \tilde T_0 (1 + \kappa \tau)\,. 
\label{T-of-t}
\ee
The constants $\gamma$ and $\beta$ are equal to
\be
\gamma^2 = \frac{3q}{(-R_c)} \left(-\frac{R_c}{\tilde T_0}\right)^{2(n+1)},\ \ \\
\beta = \gamma^2 \tilde T_0/3 = q \left(-\frac{R_c}{\tilde T_0}\right)^{2n+1}\,.
\label{gamma-beta}
\ee

\begin{figure}[ht]
\begin{center}
\includegraphics[width=.32\textwidth]{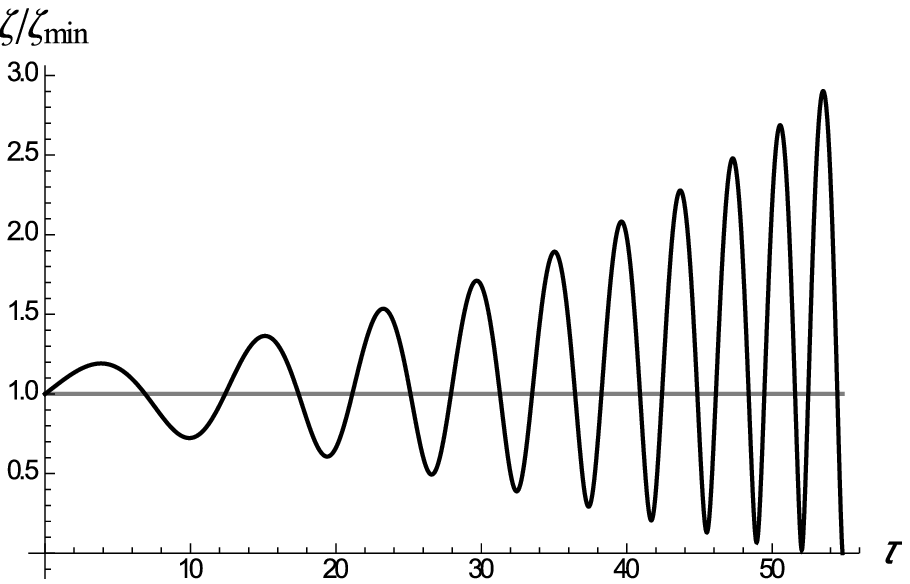} 
\includegraphics[width=.32\textwidth]{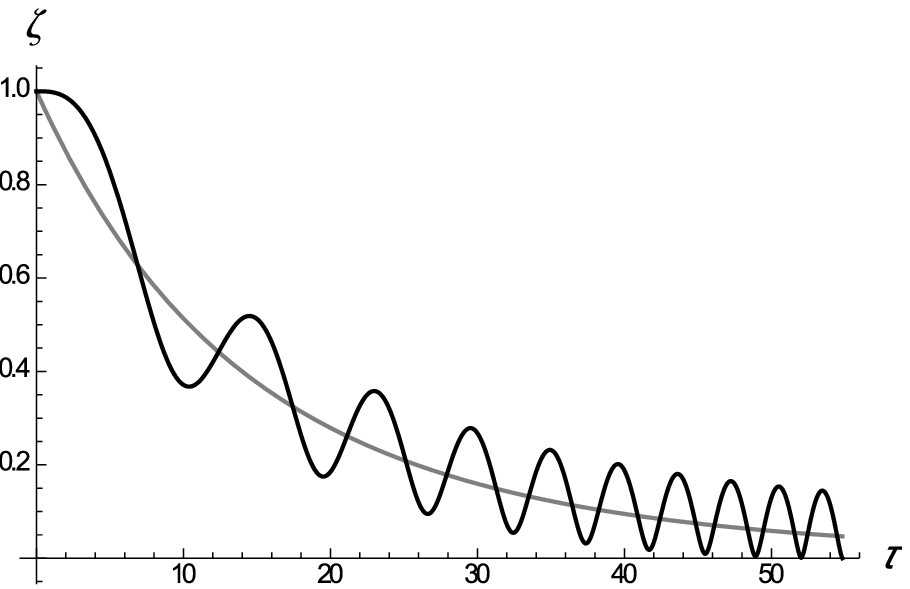} 
\includegraphics[width=.32\textwidth]{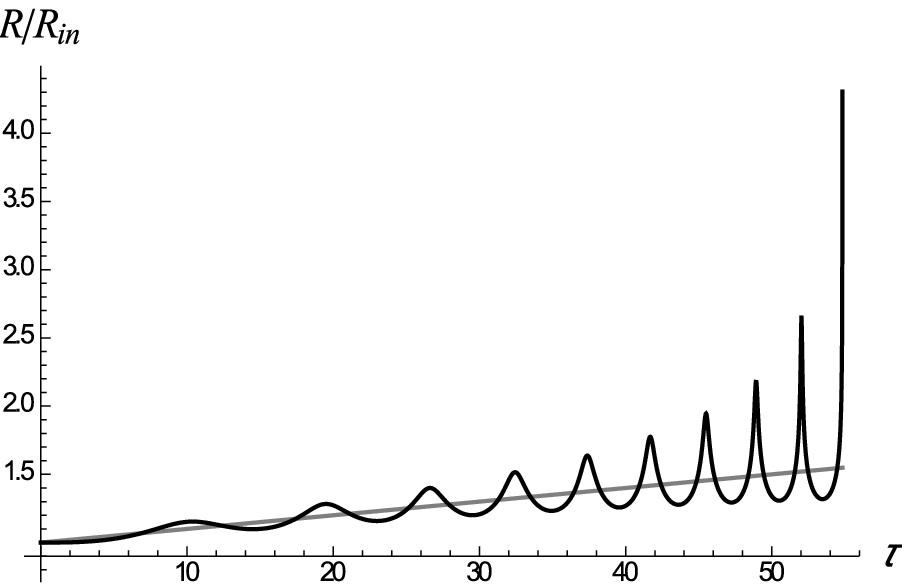}
 \caption{
Ratio $\zeta/\zeta_{min}$ (left), functions $\zeta$ and $\zeta_{min}$ (centre), and corresponding $R$ (right) for
$n=3$, $\kappa = 0.01$, $\rho_m/\rho_c = 10^5$.}
\label{f-34}
\end{center}
\end{figure}

It was shown in Ref.~\cite{Arbuzova:2010iu} that in systems with rising mass density the position of the minimum of the potential $\zeta_{min} = (1+\kappa \tau )^{-1/\nu}$ moves towards zero and so does $\zeta (\tau)$ itself, practically independently on the initial conditions, i.e. on $\zeta(0)$ and $\zeta'(0)$. The function $\zeta (\tau)$ oscillates around $\zeta_{min} (\tau)$ and at some moment it passes beyond the minimum of $U$ and reaches the point $\zeta = 0$. This value of $\zeta$ however corresponds to the 
{singularity} $R=\infty$. The system arrives to the singularity in a finite time, while the external energy density still remains finite. This conclusion is supported by the numerical calculations, shown in Fig.~
\ref{f-34}. 
For these particular figures we took $\zeta_0 = 1$ (i.e. the GR value) and $\zeta'_0= 0$. A more rigorous choice of initial conditions is discussed in section~\ref{s-init-cond}. In what follows we shift the initial time moment in such a way that $\tau=0$ corresponds to the onset of the rise of density perturbations, keeping the same notation $\tau$ for the shifted time.

\subsection{Regularization by \texorpdfstring{$R^2$}{R2}}
\label{ss-R2}

To eliminate the past and future singularities the function (\ref{eq:model-0}) was modified by adding a term proportional to the curvature squared~\cite{Starobinsky:2007hu}:
\be\label{eq:model}
F(R) = -\lambda R_c\left[1-\left(1+\frac{R^2}{R_c^2}\right)^{-n}\right]-\frac{R^2}{6m^2}\,,
\ee
where some parameters are defined below Eq.~(\ref{eq:model-0}) and $m$ has 
{ dimension of energy.} 
The additional term is relevant only at very large curvatures, because $m\gtrsim 10^5$ GeV is necessary in order to preserve 
the successful predictions of the standard BBN~\cite{Arbuzova:2011fu}.

In our works~\cite{Arbuzova:2012su,Arbuzova:2013ina} we have used the model based on Eq.~(\ref{eq:model}). 
A large $m$ implies that the stabilisation takes place at very high $R$. Though $R$ does not become infinite, it can reach huge values in systems with rising mass density. This rise normally originates after the onset of structure formation at $z\sim 10^4$ or any time later.

We are particularly interested in the regime $|R_c|\ll|R|\ll m^2$, in which $F(R)$ can be approximated by
\be\label{eq:model_approx}
F(R)\simeq -R_c\left[1-\left(\frac{R_c}{R}\right)^{2n}\right]-\frac{R^2}{6m^2}\,.
\ee
We consider a nearly-homogeneous distribution of pressureless matter, with energy/mass density rising with time but still relatively low (e.g. a gas cloud in the process of galaxy or star formation). In such a case the space derivatives can be neglected and, if the object is far from forming a black hole, the space-time metric is approximately Minkowski. Then Eq.~(\ref{trace-0}) takes the form
\be
\label{eq:trace_approx}
3\partial_t^2 F'_{R} - R - \tilde T = 0\,.
\ee
Let us introduce the dimensionless quantities\footnote{The parameter $g$ should not be confused with $\det g_\mn$.}
\be\label{eq:definitions}
\begin{gathered}
 z\equiv \frac{T(t)}{T(t_{in})}\equiv \frac{T(t)}{T_0}= \frac{\rho_m(t)}{\rho_{m0}}\,,
 \qquad y\equiv -\frac{R}{\tilde T_0}\,, \\
g\equiv \frac{\tilde T_0^{2n+2}}{6 n(-R_c)^{2n+1}m^2}= \frac{1}{6 n (m t_U)^2} \,\left( \frac{\rho_{m0}}{\rho_c}\right)^{2n+2}\,,
\qquad \tau\equiv m\sqrt g\,t\,,
\end{gathered}
\ee
where $\rho_c \approx 10^{-29} $ g/cm$^3$ is the cosmological energy density at the present time, $\rho_{m0}$ is the initial value of the mass/energy density of the system under scrutiny, and $\tilde T_0 = 8\pi \rho_{m0}/m_{Pl}^2$. Next let us introduce the new scalar field:
\be\label{eq:xi_definition}
\xi\equiv \frac{1}{2 n}\left(\frac{\tilde T_0}{R_c}\right)^{2n+1}F'_{R} = \frac{1}{y^{2n+1}}-gy\,.
\ee
In terms of this field Eq.~\eqref{eq:trace_approx} can be rewritten in the simple oscillator form:
\be
\label{eq:xi_evol}
\xi''+z-y=0\,,
\ee
where a prime denotes derivative with respect to $\tau$. The potential of the oscillator is defined as:
\be
\frac{\partial U}{\partial \xi}= z - y(\xi).
\label{U-prime}
\ee
The substitution (\ref{eq:xi_definition}) is analogous to what is done in \cite{Arbuzova:2010iu} but now $y$ cannot be analytically expressed through $\xi$ and we have to use two approximate expression for positive and negative $\xi$, see Eqs.~(\ref{xi+}, \ref{xi-}) below.

It is clear that~\eqref{eq:xi_evol} describes oscillations around $y=z$ (the ``bottom'' of the potential), 
which corresponds to the usual GR solution $R+\tilde T=0$. So we can separate the solutions into the average 
and the oscillatory part. 
For small deviations from the minimum of the potential, the solution takes the form:
\be
\xi(\tau)=\left[\frac{1}{z(\tau)^{2n+1}}-gz(\tau)\right]+\alpha(\tau)\sin F(\tau)\equiv \xi_{min}(\tau)+\xi_1(\tau)\,,
\label{eq:xi_expansion}
\ee
where
\be
F (\tau) \equiv \int^\tau_{\tau_0}d\tau'\,\Omega (\tau')\,,
\ee
and the dimensionless frequency $\Omega$ is defined as
\be\label{eq:frequency_U}
\Omega^2 = \frac{\partial^2 U}{\partial \xi^2}\,,
\ee
taken at $y=z$. From \eqref{eq:xi_evol}, we find that it is equal to
\be\label{eq:frequency_2}
\Omega^2 = -\left.\frac{\partial y}{\partial\xi}\right|_{y=z} = -\left.\frac{1}{\partial\xi/\partial y}\right|_{y=z} = \left(\frac{2n+1}{z^{2n+2}}+g\right)^{-1}\,.
\ee
The conversion into the physical frequency $\omega$ is given by
\be\label{eq:Omega}
\omega = \Omega\,m \sqrt g\,.
\ee
It is assumed that at the initial moment $\tau= \tau_0$ the function $\xi (\tau)$ sits at the minimum of the potential, otherwise we would need to add a cosine term in~\eqref{eq:xi_definition}. 
If initially $\xi (\tau_0)$ was shifted from the minimum, the oscillations would generally be stronger and the effect of particle 
production would be more pronounced.

\subsection{Scalaron Potential and Evolution \label{ss-scal-pot}}

\begin{figure}[ht]
\centering
 \includegraphics[width=.35\textwidth]{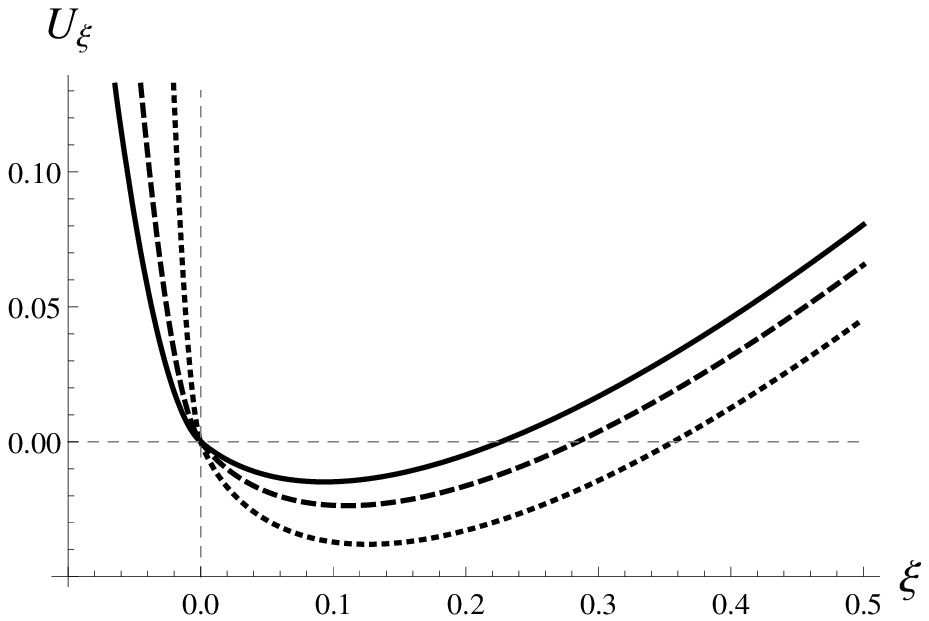}
\includegraphics[width=.35\textwidth]{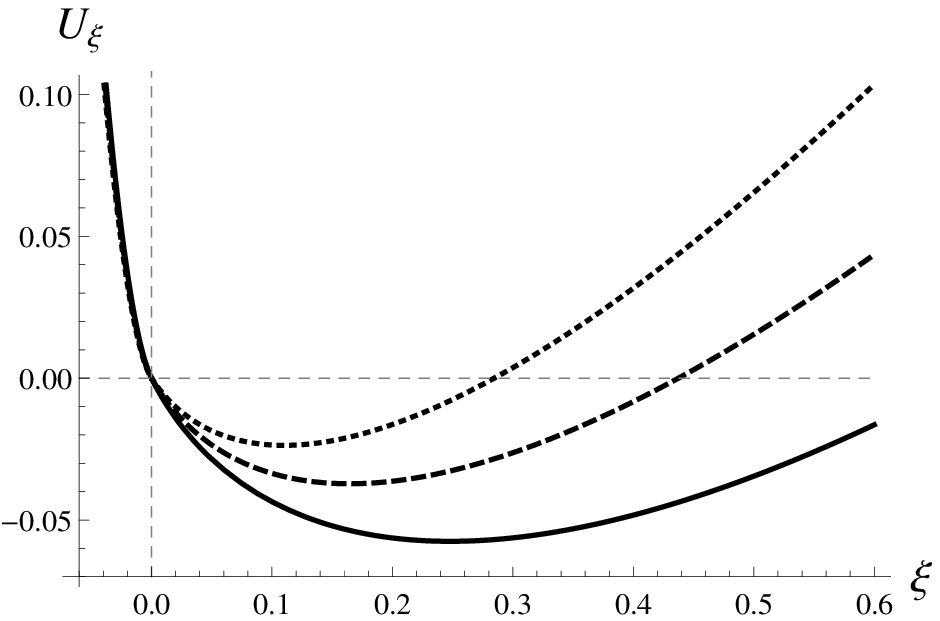}
\caption{Examples of the variation of potential~\eqref{eq:xi_potential} {as a function of $\xi$}
for different values of parameters. 
{\it Left panel} $(n=2,\,z=1.5)$ solid line: $g=0.02$, dashed line: $g=0.01$, dotted line: $g=0.002$; the part of the potential at $\xi<\xi_a$ 
($\xi_a$ is defined in Eq.~(\ref{eq:xi_expansion})) 
is increasingly steeper as $g$ decreases; the bottom of the potential also moves. {\it Right panel} $(n=2,\,g=0.01)$ solid line: $z=1.3$, 
dashed line: $z=1.4$, dotted line: $z=1.5$; the bottom of the potential moves to higher values of $U$ and lower values of $\xi$, 
as $z$ increases.}
\label{fig:potential}
\end{figure}

One cannot analytically invert Eq.~\eqref{eq:xi_definition} to find the exact expression for $U(\xi)$. However, we can find approximate expressions for $gy^{2n+2}\ll 1$ ($\xi>0$) and $gy^{2n+2}\gg 1$ ($\xi<0$). The value $\xi=0$ separates two distinct regimes, 
in each of which $\Omega$ 
has a very simple expression [see Eq.~\eqref{eq:frequency_2}] and $\xi$ is dominated by either one of the two terms in the r.h.s. 
of Eq.~\eqref{eq:xi_definition}. Hence, in those limits the relation $\xi=\xi(y)$ can be inverted giving an explicit expression for $y=y(\xi)$, and therefore the following form for the potential:
\begin{subequations}\label{eq:xi_potential}
\be\label{eq:xi_pot_theta}
U(\xi) = U_+(\xi)\Theta(\xi) + U_-(\xi)\Theta(-\xi)\,,
\ee
where
\be\label{eq:potential_+_-}
\begin{aligned}
U_+(\xi) &= z\xi - \frac{2n+1}{2n}\left[\left(\xi+g^{(2n+1)/(2n+2)}\right)^{2n/(2n+1)}-g^{2n/(2n+2)}\right]\,,\\
U_-(\xi) &= \left(z-g^{-1/(2n+2)}\right)\xi+\frac{\xi^2}{2}\,.
\end{aligned}
\ee
\end{subequations}
By construction $U$ and $\partial U/\partial\xi$ are continuous at $\xi=0$. The shape of this potential is shown in Fig.~\ref{fig:potential}.

The bottom of the potential, as it is obvious from Eq.~\eqref{U-prime}, corresponds to the GR solution $R=- \tilde T$, or $y(\xi) = z$, and its depth (for $gz^{2n+2}<1$) is
\be\label{eq:potential_bottom}
U_0(\tau) \simeq -\frac{1}{2n\,z(\tau)^{2n}}\,.
\ee

We will use a very simple form for the external energy density $z$, namely
\be\label{eq:z_linear}
\begin{aligned}
z(\tau) &= 1+\kappa(\tau-\tau_0)\\
\rho(t)&= \rho_0\left(1+\frac{t-t}{t_{contr}}\right)\\
\kappa^{-1} &\equiv m\sqrt g\,t_{contr}\,.
\end{aligned}
\ee
Here, $\kappa^{-1}$ and $t_{contr}$ are respectively the dimensionless and physical timescales of the contraction of the system; analogously, $\tau_0$ and $t_0$ are respectively the dimensionless and physical initial times, which for simplicity and without loss of generality will be taken equal to 0, though in Sec.~\ref{s-init-cond} we shift the time.

It is also useful to express the physical parameters such as $m$, the initial energy density $\rho_{m0}$, etc., in terms of their 
respective ``typical'' values. Let us define
\be\label{eq:typ_param}
 \rho_{29}\equiv \frac{\rho_{m0}}{\rho_c}\,,\quad m_5 \equiv \frac{m}{10^5\text{ GeV}}\,,\quad t_{10}\equiv \frac{t_{contr}}{10^{10}\text{ years}}\,,
\ee
where $\rho_c=10^{-29}\text{ g cm}^{-3}$ is the present (critical) energy density of the Universe. In terms of these quantities, we can 
rewrite $g$ and $\kappa$ as
\be
\begin{aligned}
g &\simeq 1.2\times 10^{-94}\,\frac{\rho_{29}^{2n+2}}{n 
\,m_5^2}\,,\\
\kappa &\simeq 1.9 \,\frac{\sqrt{n}
}{\rho_{29}^{n+1}\,t_{10}}\,.
\label{g-kappa}
\end{aligned}
\ee

The equation of motion \eqref{eq:xi_evol} for small oscillations $\xi_1$, which is defined in Eq.~(\ref{eq:xi_expansion}), can be rewritten as
\be
\label{eq:xi_evol_approx}
\xi_1''+\Omega^2\xi_1= -\xi_{min}'' \,,
\ee
{The term} $\xi_{min}''$ is proportional to $\kappa^2$, which is usually 
assumed to be small, so in first approximation it can be neglected, though an analytic solution { for constant ${ \Omega}$ 
or in the limit of large ${ \Omega}$} can be obtained with  {an account of} this term as well. 
Using expansion \eqref{eq:xi_expansion} and neglecting $\alpha''$, we obtain
\be
\label{eq:alpha_omega}
 \alpha\simeq \alpha_0\,
\sqrt\frac{{ \Omega_0}}{{ \Omega}}= \alpha_0\left(\frac{1}{z^{2n+2}}+\frac{g}{2n+1}\right)^{1/4}
{ \left({1}+\frac{g}{2n+1}\right)^{-1/4}}.
\ee
Here and in what follows sub-0 means that the corresponding quantity is taken at initial moment $\tau = \tau_0$. We impose the following initial conditions
\be
\begin{cases}
y(\tau=\tau_0)=z(\tau=\tau_0)=1\,,\\
y'(\tau=\tau_0)=y'_0\,,
\end{cases}
\label{init-y}
\ee
the first of which corresponds to GR solution at the initial moment. The initial value of the scalaron amplitude $\alpha_0$ can be expressed through $y_0'$, which we keep as a free parameter, as: 
\be 
\label{alpha-0}
{ \alpha_0 = (\kappa - y'_0) (g + 2n+1)^{3/2} }.
\ee

\begin{figure}[ht]
\centering
\includegraphics[width=.35\textwidth]{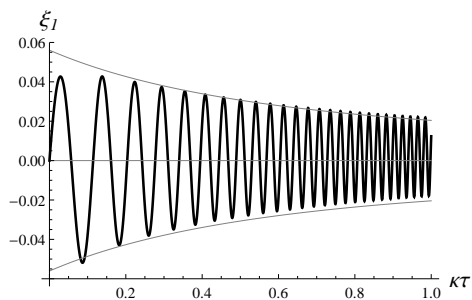}
\caption{Oscillations of $\xi_1(\tau)$, in the case $n=2$, ${ \kappa=0.01}$, ${ g=0.01}$ and initial conditions $y_0=1$, $y'_0=\kappa/2$. 
The amplitude of the oscillations is in very good agreement with the analytical estimate \eqref{eq:xi_expansion}.}
\label{fig:xi_amplitude}
\end{figure}

In our works~\cite{Arbuzova:2012su,Arbuzova:2013ina} we have chosen the initial conditions: $|y'_0| \lesssim \kappa$ and 
$|\kappa - y'_0| \sim \kappa$. Since, as we know, the evolved collapsing cloud of matter {gradually}
deviates from GR, the closeness of the 
derivative $y_0'$ to GR at the moment when the oscillating $y (\tau)$ crosses the GR value is quite unnatural. So we disregarded 
the case  $|\kappa - y'_0| \ll \kappa$.  Under these assumptions $\xi_1$ remains small, $\xi_1 \ll 1$, but non-negligible.
In this case the numerical results, shown in figure \ref{fig:xi_amplitude}, are in remarkable agreement with Eq.~\eqref{eq:xi_expansion}.

From Eq.~\eqref{alpha-0}, one is {led} to think that for $y'_0 = \kappa$ the oscillations would not be excited 
at all. This is only true at first 
order in $|y'_0 - \kappa|$, so their amplitude would be of order $\kappa^2$. It is one of the points of the criticism raised by GT. They claim 
that the system under scrutiny must start from the state determined by General Relativity (GR). It means that $ y_0 = 1$ and 
$y'_0 = \kappa$. According to the GT statement, the curvature scalar in this case would be  much smaller than it is found in our works. 
However, as we shall see below, such initial conditions could be realized only at the very onset of structure formation when 
the cosmological decrease of density stopped and the contraction started. Indeed, systems with decreasing energy density approach 
General Relativity and thus the GR initial conditions must be true. Evidently near the minimal density {its} evolution 
 is quadratic in time and density behaves as $\delta \rho \sim (\kappa \tau)^2$. However, it can be shown (see the next section) that the 
 contracting system quickly deviates from GR approaching a state in which density grows {approximately}  linearly 
 with time and ultimately realizing, at some later time, the conditions which we took in our works as the initial conditions. 
{
 {In realistic situation the density rise is closer to the exponential one in accordance with the Jeans law. It leads to
 somewhat faster excitation of the curvature oscillations and in this sense is more favorable for the discussed effect.  
}}

Our primary goal is to determine the amplitude and shape of the oscillations of $y$. Expanding $y$ as
\be\label{eq:y_expand}
y(\tau) = z(\tau) + \beta(\tau)\sin\left(\int^\tau_{\tau_0}d\tau'\,\Omega(\tau')\right)\,,
\ee
it is easy to show~\cite{Arbuzova:2013ina} that  {during} the initial oscillating phase, when $|\beta| < z$, we have
\be\label{eq:beta_evol}
|\beta(\tau)| \simeq |\alpha(\tau)|\Omega^2(\tau)\,.
\ee

\begin{figure}[!t]
\centering
 \includegraphics[width=.35\textwidth]{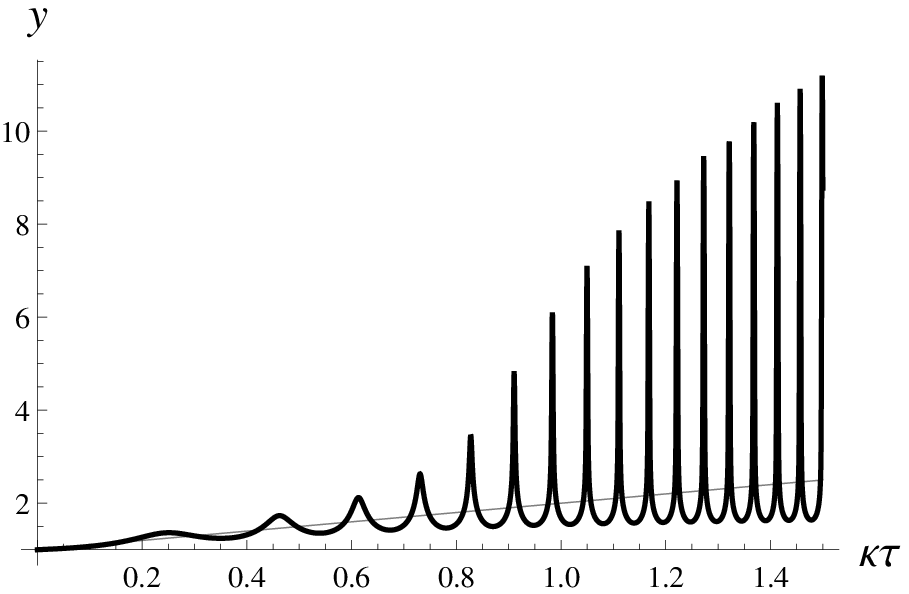}
\includegraphics[width=.35\textwidth]{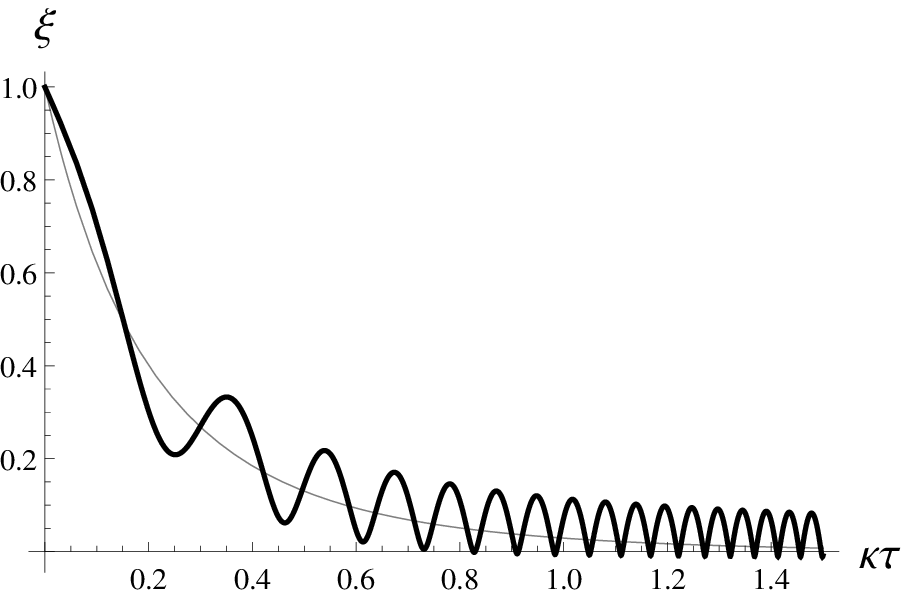}
\caption{``Spikes'' in the solutions. The results presented are for $n=2$, $g=0.001$, $\kappa=0.04$, and $y'_0=\kappa/2$. 
Note the asymmetry of the oscillations of $y$ around $y=z$ and their anharmonicity.}
\label{fig:spikes}
\end{figure}

Let us stress that, in contrast to $\xi$, the oscillations of $y$ quickly become strongly anharmonic and even for slightly negative $\xi $ the amplitude of $y$ may be very large because $y \approx -\xi /g$, according to Eq.~(\ref{eq:xi_definition}). This feature is well demonstrated by the results of numerical calculations shown in figure \ref{fig:spikes}. 

We can estimate the amplitude of the spikes analytically using the energy evolution law:
\be\label{eq:conserved}
\frac{1}{2}\,\xi'^2 + U(\xi) - \int^\tau_{\tau_1}d\tau'\frac{\partial z}{\partial\tau'}\,\xi(\tau')=\text{ const},
\ee
where $\tau_1$ is an arbitrary fixed time moment.
The last term appears because $U$ explicitly depends on time through $z$. 
If ${ \partial z/\partial \tau}$ is positive, which is the case for a contracting body, the value of $U(\xi)$ 
{would} in general grow with time. According to the assumption 
made above, $z$ linearly grows with time as $ z(\tau) = 1+\kappa\tau$, where $\kappa$ is
given by Eq.~(\ref{eq:z_linear}).
This simple law may be not accurate when $t/t_{contr}>1$, but probably the results obtained are not too far 
from the realistic case.

Let us use {this law}
when the minimum value of $\xi$ reaches zero. {Evidently in the minimum}  $\xi' = 0$. 
Since $U(0) =0$, the constant in the r.h.s. of Eq.~(\ref{eq:conserved}) turns to zero. Now let us go to larger time and neglect the 
oscillating part of $\xi$ under the integral. The 
minimum value of $\xi$ (maximum absolute value) $\xi_{min}$ of negative $\xi$ is determined by the
equation:
\be
U_- (\xi_{min}) = \kappa \int_{\tau_1}^\tau d\tau' \left[ z^{-(2n+1)} - gz\right] = 
\frac{1}{2n} \left[ \left(\frac{1}{z(\tau_1)}\right)^{2n} - \left(\frac{1}{z(\tau)}\right)^{2n} \right]
+\frac{1}{2}g \left[ {z^2(\tau_1) - z^2(\tau)} \right].
\label{U-of-xi-max}
\ee
The value of $z(\tau_1) \equiv z_1$ is found from the condition $\alpha = \xi_{max}$ i.e.
\be 
z_1^{-(2n+1)} - g z_1 = |\kappa - y' _0| \left( g +2n +1\right)^{5/4}\,\left(\frac{2n+1}{z_1^{2n+2}} + g\right)^{1/4} .
\label{z1-eq}
\ee
In the limit of small $g$, $g< 1/z_1^{2n+2}$, asymptotically $|\xi_{max}| = (g/n)^{1/2}\, z(\tau_1)^{-n}$. In the same limit, Eq.~(\ref{z1-eq}) gives:
\be
z_1 = \left[ \left( \kappa - y_0' \right)^2\,\left(2n+1 \right)^3 \right]^{-1/(3n+1)}.
\label{z1}
\ee
This finally determines 
\be {
y_{max} } = (ng)^{-1/2} z_1^{-n}.
\label{y-max}
\ee

{Let us estimate now the maximum value of $R$ or, what is essentially the same, $y_{max}$ (\ref{y-max}).
In our works we did it both ways, analytically and numerically. 
We present analytical asymptotic estimates for a small value of parameter $g$ given by Eq.~(\ref{eq:definitions}).
The numerical solutions has been found  in our works [5,6]. The agreement between numerical and analytical results is excellent.
However, numerical results can be done only for non-realistic values of parameters, e.g. small $g$ but not so tiny
as its real value, but the smaller is $g$, the better must be asymptotic solution. So we can trust both numerical
and analytical calculations. 

Analytical results, presented in this work, are copied from our papers~\cite{Arbuzova:2012su,Arbuzova:2013ina}. 
The maximum value of $R$ is equal to $R_{max} = R_{GR} y_{max}$, where $y_{max}$
is given by eq.~(\ref{y-max})
Here $R_{GR}$ is the value which the curvature would have in the GR limit, i.e. 
$R_{GR} = \rho /m_{Pl}^2$, where $\rho$ is the matter density in the contracting cloud,
 $\rho$ can be written as 
$ \rho = Q \rho_c = Q m_{Pl}^2 / t_U^2$, where  $\rho_c$ is the cosmological energy density and $Q$ is some
enhancement factor of the rise of mass density in the contracting cloud. It is reasonable to expect $Q \sim 10^5$.
Here the numerical factors of order ten are omitted. They are not important. 

Next we use expression (\ref{eq:definitions}) for $g$, Eq.~(\ref{eq:z_linear}) for $\kappa$ and 
 taking $z_1\approx (\kappa)^{-2/(3n+1)}$ in Eq.~(\ref{z1})
 we find for the maximum amplitude at the spike~$R_{max} = R_{GR} \, y_{max}$:
\be
R_{max} /m^2 = (t_U/t_{cont})^\sigma Q^{1-(n+1)(\sigma +1)} (m t_U)^{-1}
\ee
where $\sigma = 2n/(3n+1)$. The last factor is so small, that for any reasonable values of parameters $R<< m^2$ and 
thus the theory remains below its ultraviolet cut-off.
}

\section{Initial conditions}
\label{s-init-cond}

As mentioned above, the disagreement with GT is about the initial curvature velocity $y'_0$, which in turn determines the initial amplitude of the scalaron oscillations, $\alpha_0$. We assumed that $\alpha_0 \sim \mathcal O(\kappa)$, whereas GT 
argued that, because the GR solution is $y'_0 = \kappa$, $\alpha_0$ should be of order $\kappa^2$. 
Note that~\eqref{eq:alpha_omega} vanishes for $y'_0 = \kappa$ due to Eq.~\eqref{alpha-0}.

Evidently, the source term in the r.h.s. of Eq.~(\ref{eq:xi_evol_approx}) produces oscillations and hence deviations from GR 
of order $\kappa^2$ with any initial conditions, including $y'_0 = \kappa$. 
Using Eqs.\eqref{eq:frequency_2}, \eqref{eq:alpha_omega}, and 
\eqref{eq:beta_evol}, 
we observe that
\be
|\beta| \simeq |\alpha|\Omega^2 \propto \alpha_0 (2n+1)^{-1/4}\left(\frac{2n+1}{z^{2n+2}} + g\right)^{-3/4}\,,
\ee
which grows for increasing $z$ regardless of the value of $\alpha_0$, whether $\sim \kappa$ or $\sim \kappa^2$. Similarly, deviations in the 
first time derivative $y'$ from $z' = \kappa$ increase, and at some moment we will inevitably have $|y'_0 - \kappa| \sim \kappa$. Redefining 
the initial time as such moment, {we conclude that
the estimates of our works~\cite{Arbuzova:2012su,Arbuzova:2013ina} are indeed reliable.} 

Let us study {now}  the evolution of the system from the moment when the expansion of the cloud changed into 
contraction, so at $t=t_{in} = 0$ the density reaches a minimum value $\rho = \rho_0$. 
As is well known, the Tolman solution for an {over-density}, which at some moment (which we fix 
at $t=0$) decouples from the cosmological expansion and starts contracting, is:
\be\label{eq:Tolman_overdensity}
\rho(t) = \rho_0(1 + 2\pi G \rho_0 t^2)
\ee
which is clearly different from the linear dependence~\eqref{T-of-t}. However, we can show that this difference does not undermine the validity of our results. 

{ To avoid misunderstanding, expressed by one of the referees, let us stress again the following. The difference between our
way of fixing the initial conditions and that of Refs.~\cite{Gorbunov:2014fwa,Gorbunov:2014eda} is that we made an order of
magnitude guess for the value of, say, $ y' $ at a moment $t_1$, while GT suggested that a physically justified way is to fix $y'$ at 
a previous time moment $t_2$. We agree with the GT choice, but using it and numerically proceeding from $t_2$ to $t_1$, we have
found that the value of $y'$ at $t_1$  is roughly the same as it was guessed in our work. So our guessed initial condition, though taken
without rigorous justification, happens to be correct. }
Here and below the notation $t$ is used instead of $\tau$, though $y$ is a dimensionless curvature which depends on dimensionless time.

We agree that the initial conditions should be rigorously fixed at the minimum value of $\rho$. But using this prescription we have 
observed even more efficient generation of the curvature oscillations than in our more naive approach. So the 
production of cosmic rays by the mechanism of our papers~\cite{Arbuzova:2012su,Arbuzova:2013ina} is by no means suppressed 
in contrast to the claim of papers~\cite{Gorbunov:2014fwa,Gorbunov:2014eda}. For negative $t$, the density 
$\rho$ drops down  with $t \to 0$. 
At this stage $y(t)$ tends to its GR value, while $y'(t)$ tends to zero independently on the initial conditions taken at negative $t$. It is well 
known that $R(t)$ in the systems with decreasing $\rho (t)$ approaches the GR value with decreasing speed. So a study of the 
cosmological history implies the following initial conditions for Eq.~(\ref{eq:xi_evol}):
\be
y (0) = 1\,\,\, {\rm and}\,\,\, y' (0) = 0. 
\label{init-t-in}
\ee
On the other hand, in systems with rising $\rho$
the curvature $R$ runs away from GR and oscillates around the GR value with rising amplitude. 
As mentioned above, $R$ even tends to infinity in finite time if the $R^2/m^2$ term is absent. Hence at that stage the derivative $y'$
can naturally reach values of order $\kappa $ {(or even higher)}.

For a qualitative understanding of the solution we first
consider the case when the $R^2$ term is absent and the evolution of $R(t)$ or, to be more precise, of 
$\zeta (\tau) $ is {governed} by equation (\ref{eq-for-z})
with the unique analytic expression for the potential at positive $\zeta$. We remark that
when $\zeta $ reaches zero the curvature becomes infinite. The treatment of this case is technically much simpler than that with
$R^2$ present, but this simpler case well illustrates the evolution of $R(t)$ in the more realistic model determined by the $F(R)$ of Eq.~(\ref{eq:model}).

\begin{figure}[ht]
\centering
 \includegraphics[width=.35\textwidth]{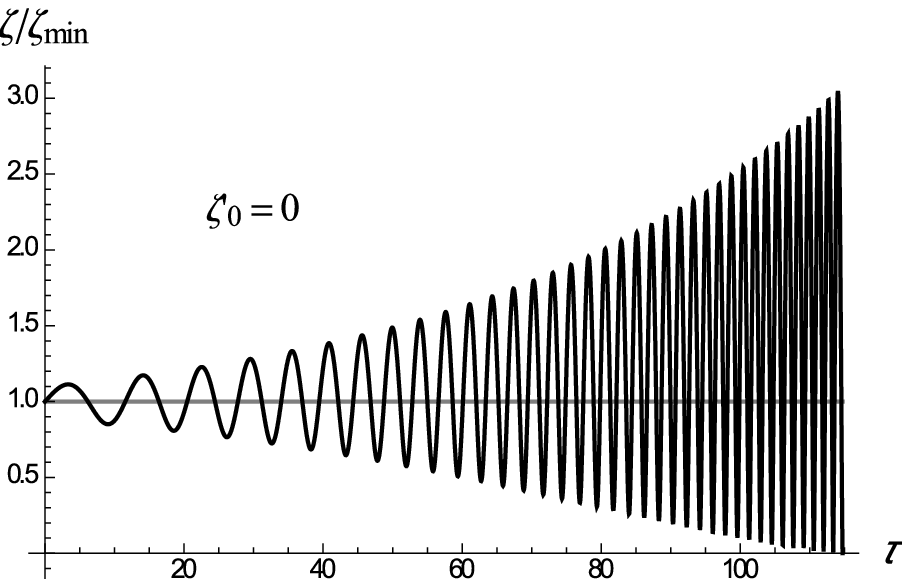}
 \includegraphics[width=.35\textwidth]{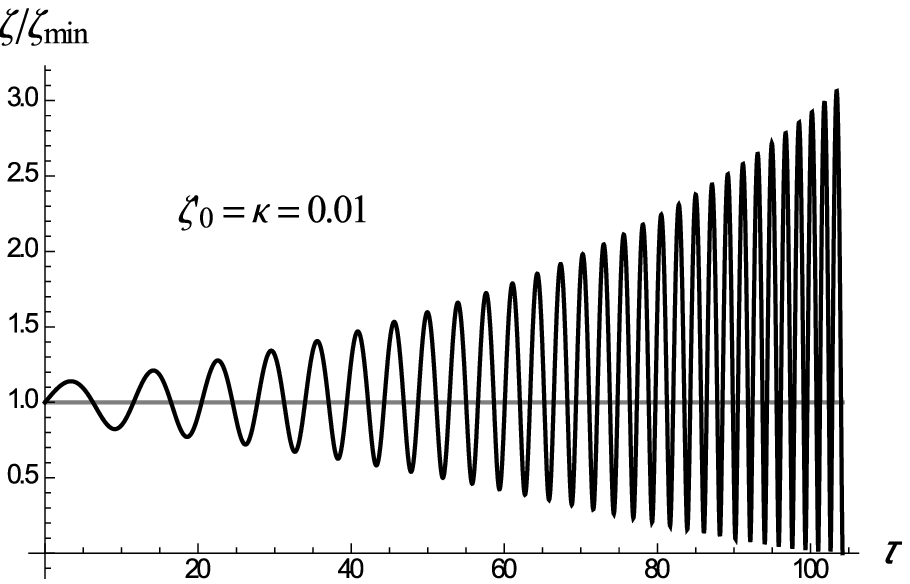}
 \includegraphics[width=.35\textwidth]{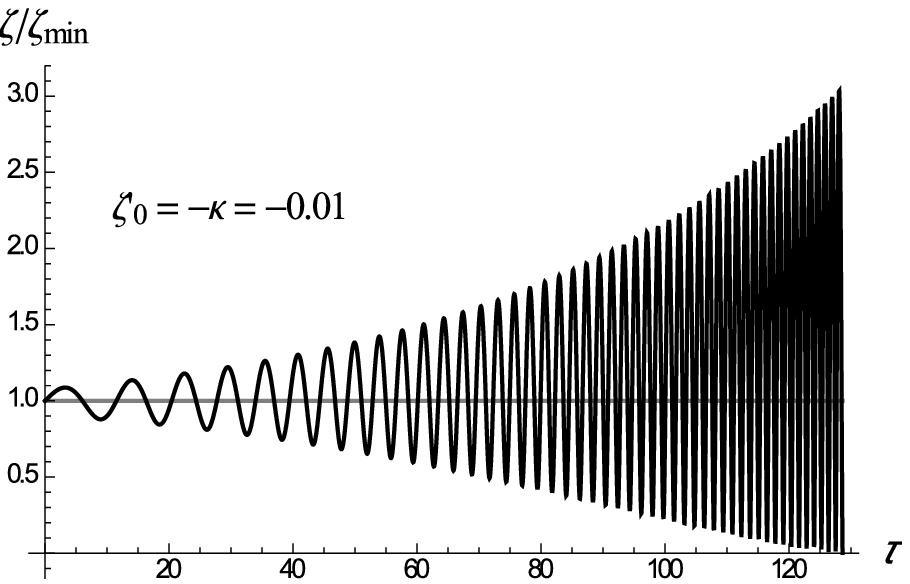}
 \includegraphics[width=.35\textwidth]{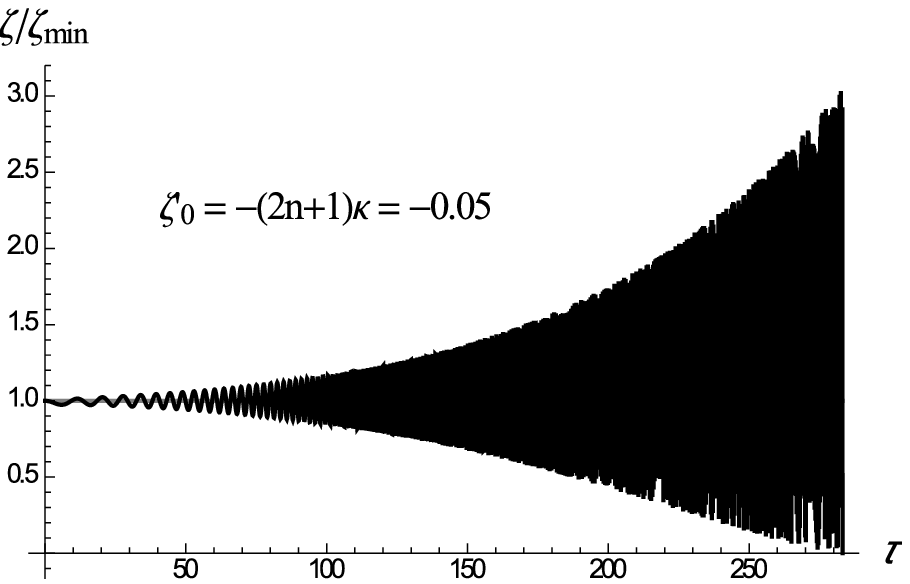}
 \caption{
{ Evolution of $\zeta/\zeta_{min}$ for $n=2$, $\kappa = 0.01$, and different initial values of the derivative 
$\zeta'_{0}$. In all the cases $\zeta$ initially sits in the minimum of the potential, $\zeta_0 =\zeta_{min}$, i.e it takes the GR value, 
while the  initial "velocities"  $\zeta'_0$ relative to motion of the minimum of the potential are different by magnitude and direction.
The least favorable case is when initially $\zeta$ moves together with the minimum of the potential, i.e.
$\zeta'_0 = -(2n+1)\kappa$.  Still in all the cases the singularity $\zeta = 0$ is reached after a time such that 
$\kappa \tau \sim \mathcal O(1)$.}
 }
\label{fig:kappa-01}
\end{figure}

{ 
We solve numerically equation (\ref{eq-for-z}) taking the initial value of $\zeta$  at the minimum of the potential, that is 
$\zeta_0= \zeta_{min} = (1 + \kappa \tau)^{-(n+1)}$ with some initial $\tau_{in}$ not necessarily equal to zero. Note that
$\zeta = \zeta_{in}$ corresponds to the GR value of curvature. We took several interesting initial values of $\zeta'_0$, i.e. of the 
"velocity" of $\zeta$ with respect to the bottom of the potential. The least favorable case for the excitation of the curvature
oscillations is $\zeta' _0 = - (2n+1) \kappa$, when initially $\zeta$ runs with the same speed as the bottom of the potential.
The results are presented in figs.~\ref{fig:kappa-01}.
}
{ We see that in all the cases $\zeta$ reaches zero roughly
speaking at $\kappa \tau \sim 1$. It means that the singularity $R =\infty$ or, better to say, high spikes of $R$
are quickly reached. }

{ Now let us turn to the physically justified initial conditions taken at the moment when the energy density given by the 
expression~ (\ref{eq:Tolman_overdensity}) acquired the minimal value, in other words, we choose
the initial conditions (\ref{init-t-in}) taken at the initial time $t_{in} = 0$. We calculate how $\zeta$ evolves starting from 
these initial conditions and check  to which
values we arrive going to the time when the density rises according to the law (\ref{T-of-t}).
In this way we find that at that time the "velocity" $\zeta'_0$ is generically of order $\kappa$ but not close to the 
special least favorable value $\zeta'_0 = -(2n+1)\kappa$. But we have seen  above that even in this case the formation of
spikes with large amplitude is  sufficiently fast to ensure efficient particle production.
}

We found no essential difference between the evolution of $ \zeta $ 
starting from many different initial values for $ \zeta'$, in particular from those  determined by the Tolman solution, 
recommended in  Refs.~\cite{Gorbunov:2014fwa,Gorbunov:2014eda}. 
Thus we conclude that the correct choice of the initial conditions at the onset of the density increase does not inhibit the 
efficiency of the curvature rise in contrast to the claim of Refs.~\cite{Gorbunov:2014fwa,Gorbunov:2014eda}. 
{In other words, the choice of the initial conditions based on the Tolmann model}
 is quite good, even more favorable 
for creation of large amplitude curvature oscillations
than the initial condition $\zeta' \sim \kappa$, as it was taken in our works~\cite{Arbuzova:2012su,Arbuzova:2013ina}.

We have done similar calculations taking $\rho(t)$ evolving according to the Tolmann solution in the theory with the $R^2$-regulator.
(Notice that we have different notations for the unknown function: $\zeta$ if $R^2$ is absent and $\xi$ if $R^2$ is present.)
The equation of motion in presence of the $R^2$-term becomes much more complicated because we have to use different 
expressions for the potential at positive and negative $\xi$. 

The equation of motion still has the simple form (\ref{eq:xi_evol}, \ref{U-prime}) but $y$ has different expressions 
through $\xi$ at positive and negative $\xi$. In the limit of small $g$ we have:
$y = \xi^{-1/(2n+1)}$, if $\xi >0$, and $\xi = -gy$, if $\xi <0$, so there are the following two equations in the two regions:
\begin{subequations}
\begin{align}
& \xi'' + (\rho_m/\rho_0) - \xi^{-1/(2n+1)} = 0, && \xi >0, \label{xi+}\\
& \xi'' + (\rho_m/\rho_0) + \xi /g = 0, && \xi <0 \label{xi-}.
\end{align}
\end{subequations}
The first equation is solved numerically, leading to a result quite close to the solution of Eq.~(\ref{eq-for-z}) for the case
without $R^2$. The initial conditions for this equation are determined from the Tolman solution in the same way as is done
above, i.e. the initial values are taken according to Eq.~ (\ref{init-t-in}). The evolution of $\xi (t)$ does not differ much from the 
previous case without the $R^2$-term as long as $\xi$ remains positive. After a while at the moment $t=t_{1}$ the almost harmonically 
oscillating $\xi$ crosses zero and we have to use equation~(\ref{xi-}). The initial values for the solution of this equation are taken as $\xi (t_{1}) = 0$, by definition, and the initial value $\xi'(t_{1})$ is determined from the numerical solution of Eq.~(\ref{xi+}). 
For $\kappa \approx 0.003$ the initial value of the derivative is calculated to be $\xi' (t_{1}) \approx -0.1$. 
For smaller $\kappa$ e.g. for 
$\kappa = 0.0003$ the derivative $\xi'(t_{1})$ is about $-0.03$. In all the cases the spikes of $R$ at negative $\xi$
are not suppressed in comparison with our simple estimates in papers~\cite{Arbuzova:2012su,Arbuzova:2013ina}.
{(We do not distinguish here between physical time $t$ and the dimensionless one, $\tau$. Hopefully it will not lead
to confusion.)}

To avoid possible misunderstandings let us stress again that the dynamical initial time is $t = 0$, while $t_{1}$ corresponds
to the initial time taken in our works~\cite{Arbuzova:2012su,Arbuzova:2013ina}. We have shown that the order of magnitude guess of the initial conditions taken in Refs.~\cite{Arbuzova:2012su,Arbuzova:2013ina} is very well supported by the calculated evolution from the
rigorously found initial conditions at $t=0$ determined  
according to the Tolman  solution.

If $g$ is much smaller than unity, Eq.~(\ref{xi-}) can be quite accurately solved analytically. To this end we rewrite it in the form:
\be 
\xi_1 '' + \xi_1/g = 0,
\label{zeta-1}
\ee
where $\xi _1= \xi + g (\rho_m/\rho_0) $.

Since the duration of the spike is very short, i.e. about $g^{1/2}$, we can neglect the time variation of $(\rho_m/\rho_0) $ during this period and the solution can be easily found. For example if we take the initial conditions as discussed above, namely $\xi(t_{1}) = 0$ and $\xi'(t_{1}) = -0.1$ the solution has the form:
\be 
y = -(\xi'_{in} /\sqrt{g}) \, \sin (t/\sqrt{g}). 
\label{y}
\ee
In this way we reconstructed the solution found in our papers~\cite{Arbuzova:2012su,Arbuzova:2013ina} in the limit of very small $g$
obtaining a spike with very high amplitude. 

We want to stress again that in our original works we fixed the initial conditions when the perturbations have already evolved and are growing roughly linearly with time. We agree that the initial conditions can be rigorously fixed at the minimum of $\rho$. Nevertheless using this prescription we have found a similar or even more efficient generation of the curvature oscillations than in our more naive approach. So the 
production of cosmic rays by the mechanism of our papers~\cite{Arbuzova:2012su,Arbuzova:2013ina} is by no means suppressed in contrast to the
statement of the papers~\cite{Gorbunov:2014fwa,Gorbunov:2014eda}.

Thus the rigorous fixing of the initial conditions based on the Tolman-type solution leads to essentially the same huge amplitude of curvature oscillations as it has been found in our works~\cite{Arbuzova:2012su,Arbuzova:2013ina} with an intuitive choice of the initial conditions. We conclude that the criticism of~\cite{Gorbunov:2014fwa,Gorbunov:2014eda} related to our choice of the initial conditions is irrelevant.

\begin{figure}[bth]
 \includegraphics[width=.35\textwidth]{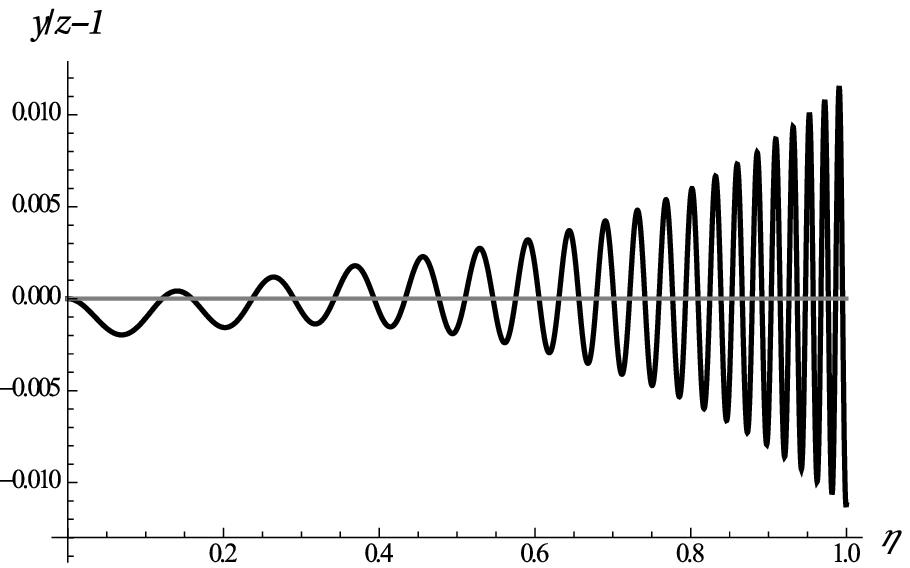}
 \includegraphics[width=.35\textwidth]{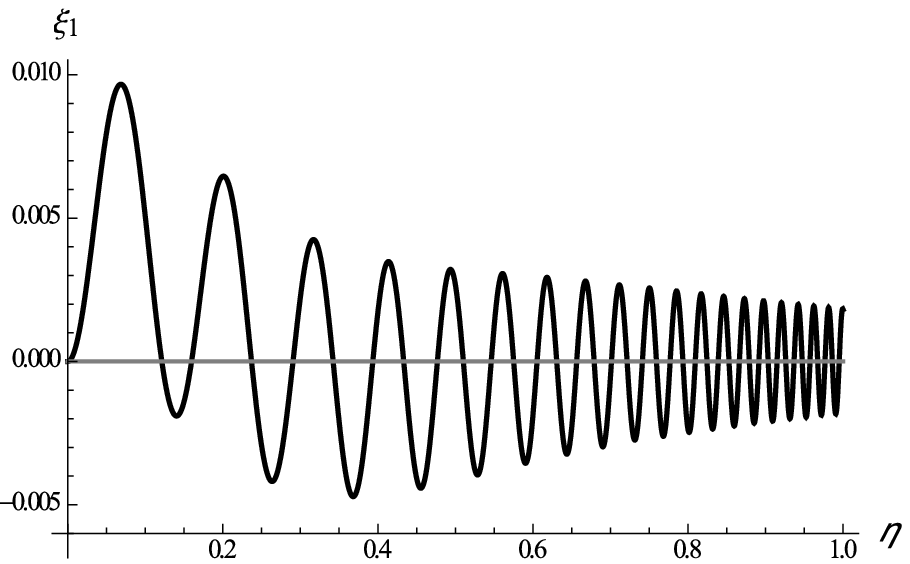}
\caption{
Behaviour of $y/z - 1$ and $\xi_1$ as functions of $\eta \equiv \sqrt{\nu}\tau$, assuming~\eqref{eq:Tolman_overdensity} or $z = 1+ \nu\tau^2$, with $\nu = 10^{-4}$, $n=2$, $g=10^{-5}$. We {choose} 
the exact GR initial conditions $(y_0,y'_0) = (1,0)$. Oscillations are still excited for both $y$ and $\xi$, qualitatively in the same way as for {the} density growing linearly in time.
}
\label{fig:oscill_Tolman}
\end{figure}

We can summarize our counter-argument on this issue as follows:
\begin{itemize}
 \item indeed the exact GR initial conditions lead to a smaller initial amplitude of the oscillations of both $y$ and $\xi$, but oscillations of $y$ grow at the same relative rate. Therefore, at some later moment we will have a more sizeable $|y'_0 - \kappa| \sim \kappa$, and our results are recovered.
 \item the fact that the oscillations of $y$ grow in time is independent of the assumed form $z(\tau)$, as long as $z' > 0$ (contraction), as shown numerically for a different functional form~\eqref{eq:Tolman_overdensity} in Fig.~\ref{fig:oscill_Tolman}.
\end{itemize}


\section{Generation of classical curvature field versus quantum production of scalarons}
\label{s-class-quan}

The second point of GT criticism refers to the probability of the scalaron production. According to their estimates~\cite{Gorbunov:2014eda}  
``one cannot expect such a large contribution of scalarons because less 
than one particle inside the horizon
may be created in the present (or recent) Universe, see~\cite{Gorbunov:2014fwa}. Scalarons created in the very
early Universe were very heavy (with mass {m}) and hence decayed to the SM particles. So
the initial conditions in~\cite{Arbuzova:2013ina} seem to be irrelevant.''  
Out of that GT concluded
that the initial number density of scalarons in the present day universe must be very small. Evidently it is not so.
The production of heavy quanta in the very early universe has nothing to do with light particle production at the present
epoch. In Ref.~\cite{Gorbunov:2014fwa} GT also considered particle production at the present day universe in the objects with 
rising energy density due to gravitational (Jeans) instability. We comment on this at the end of this Section. 

In our works~\cite{Arbuzova:2012su,Arbuzova:2013ina} we studied the generation of the classical field $R(t)$ 
{in the sufficiently late or present  day} universe after the onset of the structure formation.
 We have done explicit calculations of generation of the classical curvature field $R(t)$, which are not
related to the remnants of the early produced heavy scalaron quanta. In quantum language, which is perfectly compatible with
the classical approach, "our" quanta of scalarons, produced at the present epoch, are light.
Their number can easily be huge in the same way as low-energy radio waves correspond to a huge number of massless photons.

Clearly the heavy scalaron particles produced  in the early universe would not survive to the present day, though they may
be the source of all matter in the universe e.g. in $R^2$-inflation. 
On the other hand, our mechanism of classical generation of $R(t)$ field may result in a noticeable contribution to 
energetic cosmic rays. The classical generation of $R(t)$, of course, agrees with quantum production rate of light quanta, though
technically the calculations are more complicated.  { For example, the standard way to treat the universe heating after inflation
is to calculate particle production by  a classical oscillating scalar field (inflaton). It is equivalent to the consideration of the decay
of the inflaton condensate.}

These facts in no way undermine the validity of our results, which apply to regions of high and increasing density. As we have shown explicitly in the previous section, even starting from purely GR initial conditions, the increasing external density acts as a source term for curvature oscillations and moves solutions away from GR. In such dense, contracting systems, where $\rho_m/
\rho_c \gg 1$, certainly a contribution even much higher than $10^{-4}\rho_c$\footnote{
Which roughly corresponds to the present \textit{cosmological} contribution of the CMB. This value is mentioned by GT to ``prove'' that the initial scalaron density we assume is unphysically large.} would not affect other physical phenomena significantly, because these increase is only local, but gives much weaker contribution to the average cosmological density. 

In Ref.~\cite{Gorbunov:2014fwa} GT have considered particle production in the Einstein frame, while our calculations have been done in the  
Jordan frame. They managed to simplify the equation of motion for the scalaron mode function to a simple linear equation, Eq. (17) of their work. 
However, in this approximation all essential effects of our approach are missing. The high production probability advocated in our papers is 
induced by the non-linear term in the equation of motion and by the strong non-harmonicity of the curvature oscillations.

\section{Effects of non-harmonic oscillations of curvature}
\label{s-non-harm}

The authors of Refs.~\cite{Gorbunov:2014fwa,Gorbunov:2014eda} write: ``... production of such energetic particles in a slow and smooth process of contraction looks very surprising from any point of view...'' Another quotation from GT: ``This result [of the efficient particle production~\cite{Arbuzova:2012su,Arbuzova:2013ina}] is rather unexpected, because high frequency oscillations (i.e. heavy particles) are produced by a slow process (the structure formation) very inefficiently. Moreover, cosmological evolution naturally gives zero initial amplitude for such oscillations, which can be associated only with the scalar mode (heavy scalaron).''

In our opinion this statement is grossly incorrect. Perhaps the picture, that Gorbunov and Tokareva have in mind, originated from the assumption of harmonic oscillations of the scalaron field whose frequency and amplitude vary adiabatically. In~\cite{Arbuzova:2013ina}, we have indeed computed the predicted particle production in the initial, harmonic oscillatory phase, and found it to be negligible. However, the behaviour of $R(t)$ in the potential~(\ref{eq:xi_potential}) is { neither} necessarily ``slow'' nor ``smooth'' at all times. The fact that the statement of GT is incorrect is especially striking if the $R^2$-term is absent and $R$ tends to a singularity. The presence of $R^2$ prevents the formation of a curvature singularity but does not make the process smooth and slow. The temporal evolution of $R(t)$  is drastically different from the adiabatically slow changing quasi-harmonic oscillations. 

Our approach to particle production is very much different from that by GT. We considered the particle creation by the classical $R(t)$-field, while GT calculated the decay of the scalaron quanta, bearing in mind that the number of these quanta is negligibly small. Our results strongly contradict the GT statement: ``In all the realistic situations the scalaron production is very inefficient. Always less than one particle inside the corresponding Jeans volume is produced. Subsequent scalaron decay contribution to the cosmic ray flux is found to be in infinitesimal.'' 

Clearly the oscillating curvature scalar which is the solution of our equations~(\ref{xi+}, \ref{xi-}) corresponds to a very large number of the produced scalarons at rest. Even with the initial conditions which according to GT lead to a strong suppression of our results, the produced number density of scalarons is much larger than GT evaluated by their quantum production mechanism. 

According to GT the ``Scalarons created in the very early Universe were very heavy (with mass m) and hence decayed to the SM particles. So the initial conditions in~\cite{Arbuzova:2012su,Arbuzova:2013ina} seem to be irrelevant.'' (we repeat the quotation here to stress a different point of disagreement).
Indeed the decay rate of { scalarons in the early universe, as calculated in 
Refs.~\cite{Mamamev:1976,Zeldovich:1977,Vilenkin:1985md}  }and later  { confirmed}
 in~\cite{Arbuzova:2011fu,Gorbunov:2010bn}, is equal to:
\be 
\Gamma_R = \frac{m^3}{48 m_{Pl}^2} .
\label{Gamma-R}
\ee
With such a decay rate the expected density of scalarons should quickly drop down, if they were created in the very early universe. The similar conclusion might be valid for the classical curvature oscillations. However, this is not so in our scenario. As explained above, GT use arguments valid in a cosmological scenario but apply them to a different physical situation, namely contracting clouds with $\rho \gg \rho_c$. In our picture the curvature oscillations were induced at a late cosmological stage when the frequency of the oscillations in the quasi-harmonic regime (i.e. when $\xi $ is positive) is much lower than $m$. So the field is relatively light, and the corresponding life-time is long. 

In the spike regime the frequency is large but the duration of the spike is too short to create a relevant damping. Moreover, the expression 
for $\Gamma_R$ (\ref{Gamma-R}) is true only for harmonic oscillations~\cite{Arbuzova:2011fu}. Furthermore, a very important fact is that there is a constant energy supply for the oscillations of $R(t)$ coming from the cloud of the collapsing matter. This slow process induces the low frequency oscillations which transform their energy into high frequency modes (spikes) due to the non-linearity of the equation. Because of that the damping of the oscillations due to particle production is negligible. The validity of these statements is supported by numerical calculations.


\section{Effects of discreteness of matter}
\label{s-discrete}

As was noticed by GT there is another mechanism which might lead to the suppression of particle production by high frequency oscillations of $R$ (spikes), 
namely, the inhomogeneous distribution of the background matter consisting of separate particles. As written in Ref.~\cite{Gorbunov:2014eda}: ``The approximation that is certainly doubted is the homogeneous energy-momentum tensor for the background matter. If one considers a set of discrete particles instead of continuous medium one obtain that they produce the scalaron field like point sources. Every particle produces the field $\xi \sim \exp (-\omega r) / r $ like the usual massive scalar with mass $\omega$. If the average distance $d$ between particles is large, $d \gg 1/\omega $ then the scalaron field is strongly inhomogeneous and approximation used for the energy-momentum tensor is not valid.
In this case we expect that the energetic particle production may still take place only in a small region around the point sources, i.e. in a very small spherical regions of radius $ 1/m$. Therefore, the total number of produced particles is suppressed by a very small number $n/m^3$ where $n$ is the matter number density.''

We have already explained that the process is not smooth and slow but the problem of discrete ``constituents'' of the cosmological matter is certainly of interest. Before addressing this problem we note that this problem is not inherent to our mechanism. Indeed, the averaging problem and the matching between the small scale picture of point-like particles to a continuous perfect fluid (CDM) at larger scales is a problem of great interest even in GR. Needless to say, this problem is well beyond the scope of our original works. Therefore, as in all previous works on $f(R)$ gravity before us, we assume that a matter distribution homogeneous \textit{on average} is for all practical purposes equivalent to a truly homogeneous density distribution.

{ {Let us stress again that similar or exactly such problem exists in the standard cosmology. The 
universe is described in the ideal liquid approximation with the average density distribution. This is true in the 
case of linear equations, when the averaging of the equation leads to the equation for the average values. However,
this is not so  for the nonlinear GR equations.
Nevertheless no other work found/assumed any difference between "strictly homogeneous" and "homogeneous on average" in the
models of modified gravity (and there are very good reasons that the equality holds rather well in GR, probably thanks essentially to 
Gauss's Law). So, unless GT show explicitly that this equivalence is wrong, we see no reason why we should not use a fluid 
description for Dark Matter.}}

Moreover, the average distance between the matter particles is model dependent and might be sufficiently short  
{ or they may be even overlapping in the quantum mechanical sense. This allows to avoid the problem indicated by GT.
As a possible counter-example we can mention the case of Bose-condensate dark matter,  composed of light (e.g. axions) or 
heavy scalars, where the particle constituting dark matter forms a continuous, nearly homogeneous condensate.

We can also consider normal dark matter particles or neutral atoms of the usual baryonic matter and recall that they are not point-like 
particles but quantum mechanical wave packets. The wave packet size is determined by the coherence length of the particle 
propagation in cosmic gas/plasma so the packets would easily overlap and the quantum mechanical density matrix of the cosmological 
medium would be sufficiently homogeneous.

{ The size of the wave packet of a particle produced by the decay of a pure state of some parent particle is equal to the parent
particle lifetime. For example the size of the wave packet of a muon created in the pion decay $\pi \rightarrow \mu \nu$ is
macroscopically large, $L = \tau= 2.6\cdot 10^{-8}$ s $\approx 70$ m.  However, it is realistically much shorter because of the
breaking of coherence of the pion wave packet, see e.g.~\cite{Dolgov:2002wy}. For example, if the pion is in a thermal bath, then the coherence is maintained for no longer than a typical time $\delta t \sim v/l_{free}$,
where $l_{free}$ is the mean-free path of the source particle and $v$ is the relative (M{\"o}eller) velocity of the colliding 
particles. Bearing in mind applications to the cosmological plasma dominated by relativistic matter we take $v = 1$. So it seems natural to 
suggest that the wave packet size in thermal gas is determined by the particle mean free path in the gas, where this particle is created and
propagates  (see e.g.~\cite{Giunti:1997wq, Torres:2007zz} and references therein). { So the packet size is:}
\be
L \sim {l_{free}} = \frac{1} {\sigma_{tot} n }\,, 
\label{pack-size}
\ee
where  $\sigma_{tot}$ and $n$ are respectively
the total cross-section of the particle interactions with the cosmic plasma  and the plasma density. 
If we take some heavy matter particles, their decoupling from the rest of the plasma, or better to say, gas
takes place when:
\be
\sigma_{tot} v n \sim H,
\label{decpl}
\ee
where $H \sim T^2/m_{Pl}$ is the Hubble parameter, $T$ is the plasma temperature
and  $m_{Pl}$ is the Planck mass.  Assuming that the plasma is dominated by relativistic matter we take $v =1$. 
The plasma density is $n \sim T^3$ and the interaction cross-section is
$\sigma_{tot} \sim \alpha^2/ m_{DM}^2$, where $m_{DM}$ is  the  mass of the dark matter particle and $\alpha$ is the
coupling constant, which is typically small, $\alpha \lesssim 10^{-2}$. Correspondingly we find for the decoupling temperature:
\be 
T_d \sim \alpha^{-2} m_{DM}^2 / m_{Pl}.
\label{T-d}
\ee
Hence the wave packet size at the decoupling would be:
\be
L_d \sim 1/H_d \sim \frac{\alpha^4 m_{Pl}^3}{m_{DM}^4}
\label{L-d} .
\ee
The wave packet would be stretched out by the cosmological expansion and by the quantum mechanical packet
spreading. The cosmological expansion gives the factor $z = T_d/T_0 = m_{DM}^2/(\alpha^2 T_0 m_{Pl})$,
where $T_0$ is the CMB temperature at the present day, 
so the wave packet size would become:
\be 
L_0 \sim \frac{m_{Pl}^2}{\alpha^2 T_0 (z+1) m_{DM}^2}.
\label{L-0}
\ee
The cosmological number density of dark matter particles is equal to 
\be
n_{DM} \approx (z+1)^3 ({\rm keV /m_{DM}) /cm^3},
\label{n-DM}
\ee
where $z$ is the redshift with respect to the present time. Correspondingly the average distance between the DM-particles
is
\be
d_{DM} = (m_{DM} /\rm{keV})^{1/3} (z+1)^{-1} \,\,{\rm cm}. 
\label{d-DM}
\ee
So
\be
\frac{d_{DM} }{ L_0} = \left(\frac{m_{DM}}{m_{Pl}} \right)^2 \left(\frac{m_{DM}}{{\rm keV}} \right)^{1/3}. 
\label{l-DM-to-L0}
\ee
Evidently this ratio is very small for any reasonable $m_{DM}$ and thus the dark matter particles in cosmology are not 
well separated rigid "stones" but overlapping quantum states with very homogeneous density matrix. However, if the dark
matter particles are primordial black holes, then they would not overlap.  

Now let us show that  the GT claim that the distance between particles $d$ is much larger than the interaction scale in the harmonic regime, 
$d \gg 1/\omega$ (see the beginning of this section), is grossly overestimated. Indeed, according to Ref.~\cite{Arbuzova:2015uga}:
\be 
\omega = \left[ \frac{1}{m^2} + \frac{1}{R_c}\,\left(\frac{R_c}{R}\right)^{2n+2} \right]^{-1/2}.
\label{omega-estim}
\ee
If we take for example $R/R_c = 10^5$, $n = 3$, $R_c = 1/t_U^2$, and assume that the $1/m^2$ term can be neglected, the screening
length would be 
\be
1/\omega = t_U (R_c/R)^{n+1} \approx 3 \cdot 10^{18 - 20} \,{\rm sec} \approx 10^9 \,{\rm cm}
\label{1-over-omega}
\ee
If we take a cloud of size 1 kpc and mass of, say, $10^5 M_\odot \approx 10^{72}$ { GeV}, then the average distance between
the protons would be $\sim 10^{-3}$ cm. If the cloud consists of dark matter particles the distance would scale as the 
cubic root of the ratio of the masses of DM particles to that of protons.

The same result can be obtained if we consider the solution of equation~(\ref{eq-for-w}) in the static spherically symmetric case, 
which in the limit of $m^2 \gg R \gg R_c$ takes the form:
\be 
w''_{rr} + \frac{2}{r} w'_r + \frac{ (2n\lambda)^\nu (-R_c) }{3 w^\nu} - \frac{\tilde T}{3} = 0.
\label{w-rr}
\ee
Let us consider a sphere with homogeneous matter density $\tilde T = const$. In such a sphere the static solution of the equation of
motion coincides with the usual GR one, i.e
$R = - \tilde T$. Outside the sphere, where $\tilde T =0$, the solution should tend to the cosmological one $ R \sim R_c$. 
In terms of $w$, this means that inside the sphere $w \sim (R_c/R)^{2n +1} \ll 1$, while outside at large distances from the sphere $ w $ tends to values of order unity, i.e. $R \sim R_c$. One should keep in mind that when $w $ becomes close to unity, the equation of motion changes because Eq.~(\ref{w-rr}) is true only in the limit of $w \ll 1$.

Since according to Eq.~(\ref{w-rr}) $w''_{rr}$ is negative, the rise from $w \ll 1$ to $w\sim 1$ can be achieved if 
the boundary value of the derivative $w'_r$ is sufficiently large. The characteristic distance $r_{ch}$
from the sphere to the point where $w$ noticeably increases can be estimated as follows. In the limit we consider, the term $R^2/m^2$  is not essential, so the only dimensional parameter is $R_c$. Hence from Eq.~(\ref{w-rr}) it follows that 
\be
r_{ch} \sim w'_r w^\nu t_U^2\,.
\label{r-ch}
\ee 
Since $w'_r \sim w/r$, we find $r_{ch}^2 \sim w^{\nu+1} t_U^2$. Keeping in mind that $w^{\nu +1} = (\rho_c /\rho_m)^{(2n+2)/(2n+1)} $, 
we can conclude that
\be 
r_{ch} \sim t_U (\rho_c/ \rho_m)^{n+1}\,.
\label{r-ch-2}
\ee
This coincides with the estimate (\ref{1-over-omega}) obtained above, because the ratio of the matter density $\rho_m$ to $\rho_c$ inside the sphere is the 
same as the ratio of the curvature scalars, $\rho_m/\rho_c = R /R_c$.

One more comment is in order here. As we have shown, the average distance between quantum particles can be much smaller than the
damping distance of the curvature scalar. By definition quantum particles are those for which the Compton wavelength,
$l_C = 1/m$, where $m$ is the particle mass, is of the order of the particle size, $l_{part}$. For classical particle the relation is the 
opposite, $l_C \ll l_{part}$. Because of possible small number density of classical DM particles, such as e.g. primordial black holes, 
the distance between them, $r_{ch}$, might indeed be larger than $1/\omega$ and the GT arguments may seem to be applicable. However, as
we have mentioned at the beginning (in the second paragraph) of this section, this is a general problem of averaging of a non-linear
function inherent to canonical GR. Furthermore, the estimate by GT presented above is valid only in the quasi-static approximation, which is strongly 
broken in the spike region. The rising (spiky) solution is quickly oscillating with time, thus the field is not Yukawa-like, namely static and exponentially decreasing with distance from the source, but closer to a standing scalar wave.
}}

\section{Conclusion}
\label{s-conclusion}

To summarize, the arguments against our works~\cite{Arbuzova:2012su,Arbuzova:2013ina} by GT~\cite{Gorbunov:2014fwa,Gorbunov:2014eda} can be separated into the following  statements:
\begin{enumerate}
\item ``The choice of the initial conditions for the equations of motion~(\ref{trace-0}, \ref{eq:xi_evol}) governing the evolution of the curvature scalar $R$ is wrong''. 
{ 
We disagree that our initial conditions are wrong. We made this choice as a simple estimate by order of magnitude. 
 Numerical calculations with the initial time rigorously taken at
the onset of the structure formation justify our guess for the "initial" conditions at later time. So
 we arrive to the same conclusion, as is argued in our pubished works, 
 of the strong curvature oscillations and to unsuppressed particle  production in contrast to the GT statement.  
 We have also shown that the curvature oscillations with growing amplitude are produced  practically regardless of the 
 initial condition and of the functional form of the  rising density, $\rho (t)$.}

\item
Another critical point by GT is that any realistic matter distribution is discrete and the distance between the particles is much larger than the scale $\ell \sim \omega^{-1}$ at which $R$ drops down outside the source particles. First of all, the argument by GT relies on the assumption that we cannot treat a distribution of particles homogeneous on average as a homogeneous fluid, and moreover that the scalaron field decays following a Yukawa profile outside the particles. The first issue is indeed very interesting and relevant even in GR, but we opted to follow all literature before us in considering that one can indeed approximate a distribution homogeneous on average as truly homogeneous. As for the second assumption, it is strictly true only in the quasi-static approximation which is clearly violated in our quickly oscillating, spiky solutions, so we do not believe it to be a crucial point against our results.

Moreover, the particles constituting the system might not be classical but quantum particles whose wave packet size is much larger than $\ell$. There are also scenarios with Bose-Einstein condensate dark matter for which the density distribution is continuous and the GT arguments are not applicable.

\item
There is an essential difference between our works and the critical scenario by GT. We solved the classical equation of motion governing the evolution of practically massless field $R(t)$ and calculated particle production by this oscillating field. Naturally in a quantum language the number of the $R$-quanta in this classical state can be large. Such situation is realised for instance in the process of emission of an electromagnetic wave, which carries an almost infinite number of photons. The solution of the classical equations is straightforward and simple, and the number of the produced quanta is huge for any initial conditions, ours and those taken by GT.


\item
One more critical argument by GT is the statement that an adiabatic slow process cannot lead to an efficient creation of the classical field. However, this reasoning might be applied only to slow harmonic oscillations, while in our case the oscillations are strongly non-harmonic and non-adiabatic.

\end{enumerate}

\acknowledgments
We appreciate our discussions with D. Gorbunov and A. Tokareva and though we did not come to a consensus about the strength of the cosmic ray production, our interaction was helpful for a better understanding of possible problems with the mechanism of our works~\cite{Arbuzova:2012su,Arbuzova:2013ina}, especially of the problems with the choice of the initial conditions. The work of EA and AD was supported by the RSF Grant N 16-12-10037. LR is supported by ESIF and MEYS (Project CoGraDS -CZ.02.1.01/0.0/0.0/15\_003/0000437).


\end{document}